\documentclass[aps,prl,twocolumn,superscriptaddress,longbibliography]{revtex4-2}

\usepackage[colorlinks=true,citecolor=blue]{hyperref} % hyperreferences
\usepackage{graphicx}% Include figure files
\usepackage{dcolumn}% Align table columns on decimal point
\usepackage{bm}% bold math

\begin{document}

\title{Andreev Reflection and Klein Tunneling in High-Temperature Superconductor/Graphene Junctions}

\author{Sharadh Jois}
 \email{joiss@sunypoly.edu}
\affiliation{%
 College of Nanoscale Science and Engineering,
 SUNY Polytechnic Institute, Albany, NY 12203, USA
}%

\author{Jose L. Lado}
    \email{jose.lado@aalto.fi}
\affiliation{%
 Department of Applied Physics,
 Aalto University, 00076 Aalto, Espoo, Finland
}%

\author{Genda Gu}
\affiliation{%
 Condensed Matter Physics and Materials Science Department,
 Brookhaven National Laboratory, Upton, NY 11973, USA
 }%

\author{Qiang Li}
\affiliation{%
 Condensed Matter Physics and Materials Science Department,
 Brookhaven National Laboratory, Upton, NY 11973, USA
 }%
\affiliation{%
Department of Physics and Astronomy, 
Stony Brook University, Stony Brook, New York 11794, USA
}%

\author{Ji Ung Lee}
    \email{leej5@sunypoly.edu}
\affiliation{%
 College of Nanoscale Science and Engineering,
 SUNY Polytechnic Institute, Albany, NY 12203, USA
}%
\date{\today}% It is always \today, today,

\begin{abstract}
Scattering processes in quantum materials emerge as resonances in electronic transport, including confined modes, Andreev states, and Yu-Shiba-Rusinov states. However, in most instances, these resonances are driven by a single scattering mechanism. Here we show the appearance of resonances due to the combination of two simultaneous scattering mechanisms, one from superconductivity and the other from graphene p-n junctions. These resonances stem from Andreev reflection and Klein tunneling that occur at two different interfaces of a hole-doped region of graphene formed at the boundary with superconducting graphene due to proximity effects from Bi$_2$Sr$_2$Ca$_1$Cu$_2$O$_{8+\delta}$. The resonances persist with gating the from p$^+$-p and p-n configurations. The suppression of the oscillation amplitude above the bias energy which is comparable to the induced superconducting gap indicates the contribution from Andreev reflection. Our experimental measurements are supported by quantum transport calculations in such interfaces, leading to analogous resonances. Our results put forward a hybrid scattering mechanism in graphene/high-temperature superconductor heterojunctions of potential impact for graphene-based Josephson junctions.

\end{abstract}

 %Use showkeys class option if keyword display desired

\maketitle

Transport studies across superconductor/metal junctions have led to the discoveries of resonant interference mechanisms \cite{Rowell1966} and Andreev bound states \cite{Lofwander2001,Dirks2011}. Graphene is a two-dimensional Dirac semi-metal whose carrier density can be controlled by electrostatic gating \cite{Novoselov2005}. Bi$_2$Sr$_2$Ca$_1$Cu$_2$O$_{8+\delta}$ (BSCCO) is a layered high-temperature superconductor that has a pseudogap of 40 meV \cite{Hamidian2016,Yu2019,Liao2018}. Proximity-induced superconductivity in graphene and topological insulators from BSCCO has shown to induce a superconducting gap of 15-20 meV \cite{Wu2019,Zareapour2012,Zareapour2017}. Characterizing the transport across a BSCCO/graphene junction in response to bias, gating, and temperature can reveal new scattering mechanisms which are important for fundamental understanding and enabling applications such as high-temperature Josephson junctions. In addition, nodal superconductivity in graphene could host Majorana zero modes \cite{DiBernardo2017,Black-Schaffer2014}, which have major potential applications for fault-tolerant quantum computing.

Here, we show that two scattering mechanisms, Andreev refection (AR), associated with superconductor/metal interfaces, and Klein tunneling (KT) \cite{Katsnelson2006,Cheianov2006,Reijnders2013}, associated with Dirac fermions, occur together \cite{Beenakker2008,Burset2009} in our BSCCO/graphene junctions. They give rise to Fabry-P{\'{e}}rot resonances that attest that pristine junctions can be formed with long carrier coherence. Our results are supported by calculations of the spectral functions for graphene with proximity-induced nodal superconductivity and the emergent resonant modes in the junction.

KT is one of the most unique properties in graphene and manifests as an asymmetry in the conductance between p-p and p-n junctions \cite{Huard2007,Stander2009,Sutar2012}. The potential walls in such devices act as partially reflective interfaces which confine carriers and lead to Fabry-P{\'{e}}rot resonances \cite{Shytov2008,Young2009,Reijnders2013,Rickhaus2013,Handschin2017} in p-n-p junctions. AR is the process in which an electron from the normal metal enters the superconductor, creates a Cooper pair, and reflects as a hole. AR in graphene has a specular case at low energy \cite{Beenakker:2006}. Our work presents a unique consequence of resonant states forming from KT and AR occurring at two different interfaces.

\begin{figure}[htp]
    \centering
    \includegraphics[width=\columnwidth]{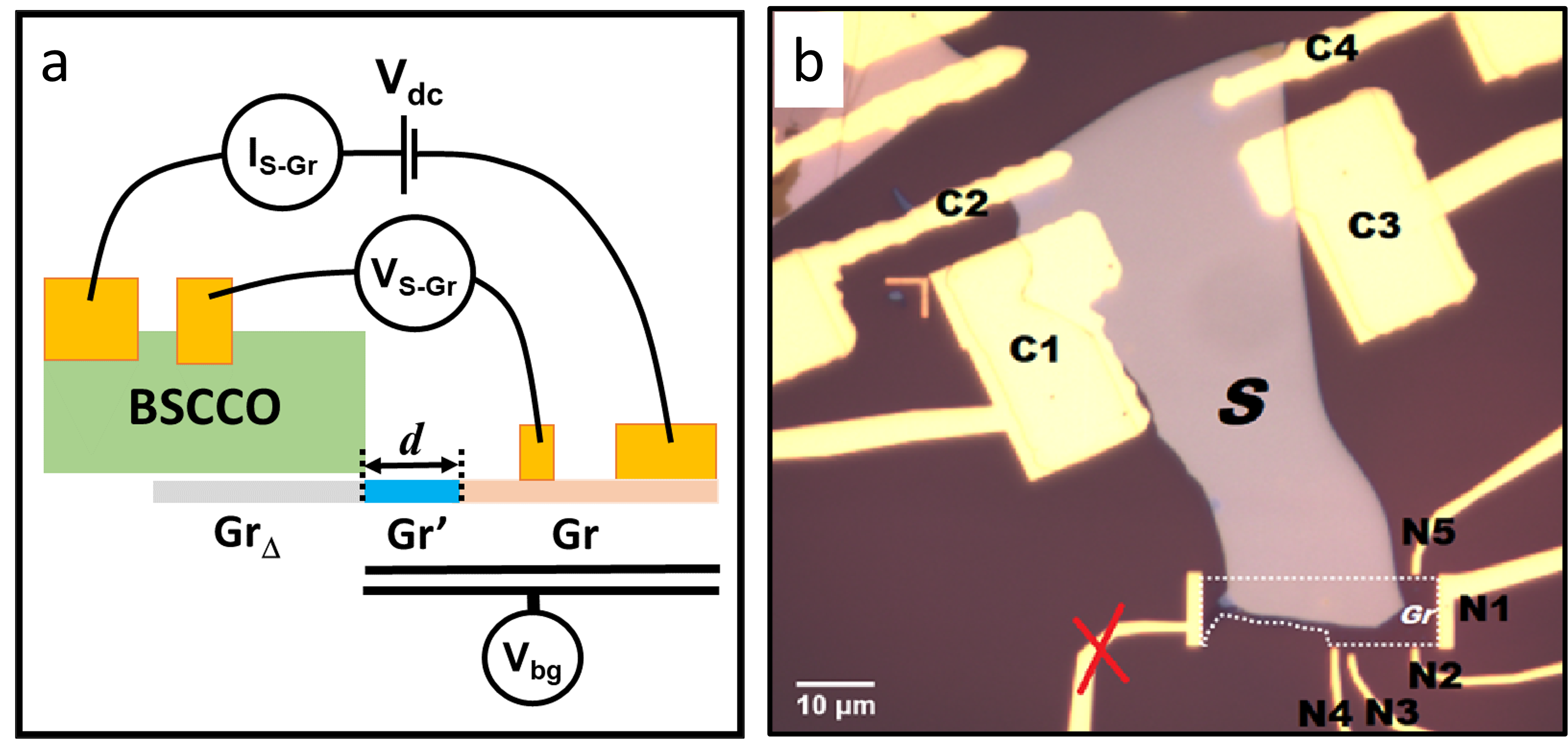}
    \caption{(a) Schematic cross-section of our device. Gr$_\Delta$ is superconducting graphene under BSCCO, Gr' is a p-doped graphene and Gr is native graphene. $d$ is the length of Gr'. The dotted lines at either ends of Gr' denote the interfaces. (b) Optical micrograph of device-A from table-\ref{tab:table1}. S is shorthand for BSCCO. Dotted white outline tags graphene. The red X and N1 to N5 are contacts to graphene. C1 to C4 are contacts to BSCCO. Scale bar is 10 $\mu$m. 
    }
    \label{fig:BSCCO-Gr junction}
\end{figure}

\begin{figure*}[htp]
    \centering
    \includegraphics[width=17.8cm]{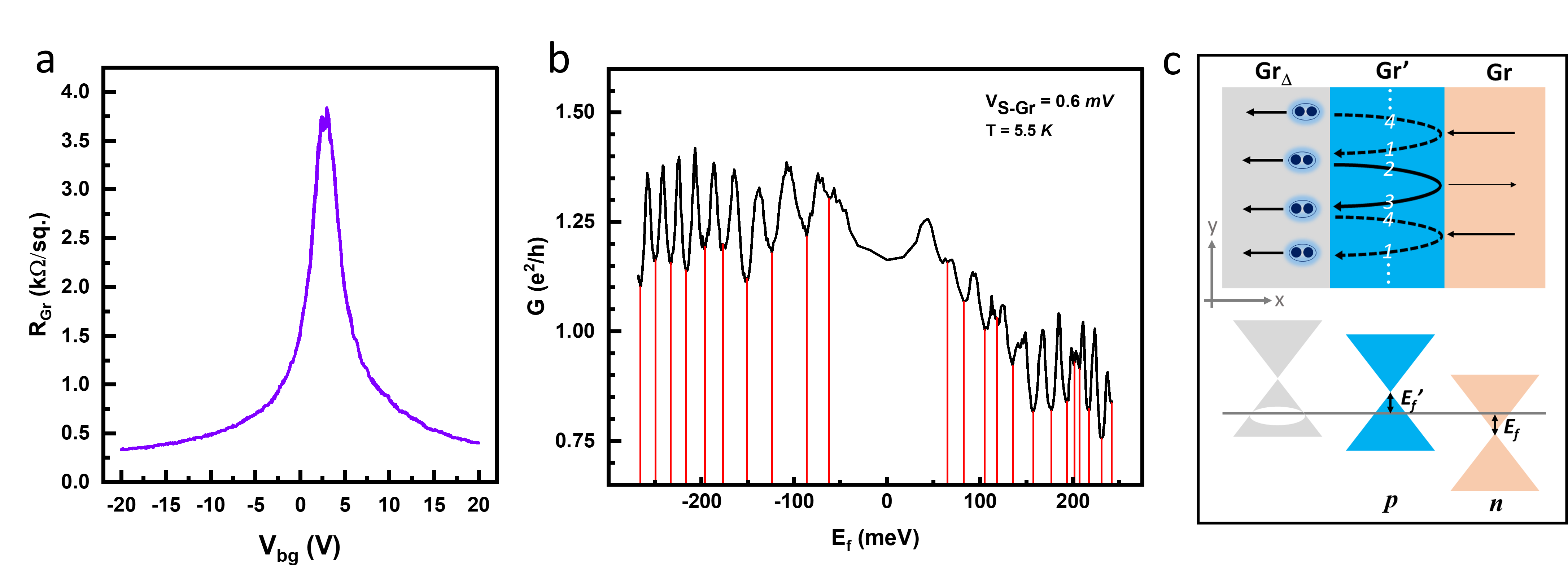}
    \caption{(a) Sheet resistance of typical bare graphene in our devices. (b) Conductance of BSCCO-Gr (device-A) junction in response to sweeping the Fermi-energy $E_f$. The red lines indicate the position of the troughs. (c) Top - A p-n junction is formed in Gr'-Gr by electrostatic gating. The classical trajectory of the holes and electrons is depicted by dashed and solid lines respectively. The number labels serve to count the trips.}
    \label{fig:Oscillations}
\end{figure*}

Three distinct regions of graphene form in our BSCCO/Gr junction. The regions are indicated as Gr$_{\Delta}$, Gr' and Gr in FIG.\ref{fig:BSCCO-Gr junction}-a. Gr$_{\Delta}$ is the portion of graphene under BSCCO that has a nodal proximity-induced superconducting gap ($\Delta_i$). BSCCO is a hole-doped superconductor with a high carrier density of $10^{15}$ cm$^{-2}$ \cite{Sterpetti2017}. BSCCO in contact with graphene of lower carrier density ($10^{12}$ cm$^{-2}$) transfers charge to create a p-doped Gr' region of length $d$. A similar charge transfer process occurs in metal/graphene junctions \cite{Mueller2009}. The dotted line between the Gr$_{\Delta}$ and Gr' regions indicates the interface which is responsible for AR. At the other end of the Gr' region is another dotted line to indicate the second interface with native graphene Gr which is responsible for KT. Therefore, the superconducting Gr$_{\Delta}$ region is connected to a p-n junction formed in the Gr'-Gr regions by a gate. Modulation of the gate results in an asymmetric conductance as the doping in the Gr'-Gr regions changes from p$^+$-p to p-n as we show below, beyond what was previously reported in YBCO/graphene junctions\cite{Perconte2018, Perconte2020,Perconte2022}.  

To fabricate our devices, we exfoliate graphite on 90 nm of silicon dioxide (SiO$_2$), which is thermally grown on a highly doped silicon wafer that is used as a global gate. Optical micrographs and Raman spectra are collected to identify monolayer graphene flakes. The graphene/substrate is then annealed at 300 \textdegree C for 3 hours in a tube furnace with a continuous flow of argon (50 sccm) and hydrogen (100 sccm) to remove residual hydrocarbons. We exfoliate high-quality BSCCO in an inert nitrogen-filled glovebox with $<0.5$ ppm water vapor and $<0.1$ ppm oxygen on 90 nm SiO$_2$ substrates. We find BSCCO flakes thicker than 25 nm and use a polypropylene carbonate / PDMS / glass slide stamp to pick it up at 45 \textdegree C. BSCCO is brought into contact with graphene at 40 \textdegree C and is finally released at 65 \textdegree C. To make contacts, we first use electron beam lithography to make ohmic contacts to graphene by evaporating Ti (3 nm) and Au (50 nm). Next, we use photolithography to make ohmic contacts to BSCCO by evaporating Ag (10 nm) and Au (40 nm). We immediately load the completed devices in a cryostat for measurements. We obtain $T_C\sim85$ K for BSCCO in our devices. An optical image of the device-A discussed here is shown in FIG.\ref{fig:BSCCO-Gr junction}-b. The junction resistance ($R\times W$) for this device is 628 $k\Omega \mu m$. Parameters for all devices can be found in Table-\ref{tab:table1}. 

\begin{figure}[htp]
    \centering
    \includegraphics[width=8.6cm]{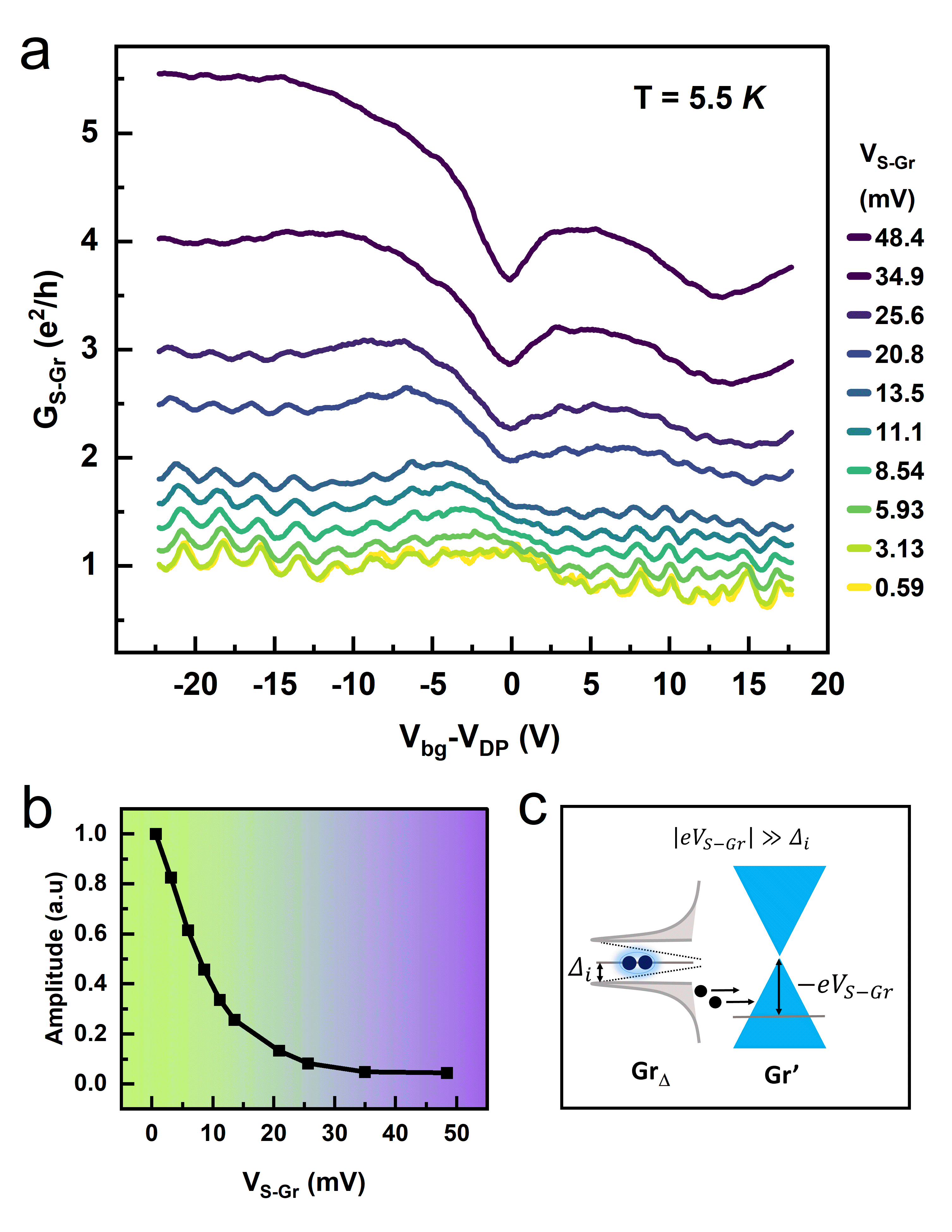}
    \caption{(a) Conductance of the S-Gr junction as a function of the back-gate voltage ($V_{bg}$) measured at different V$_{S-Gr}$, corrected for the position of the Dirac point V$_{DP}$ of Gr at +2.3 V  (b) Amplitude of the oscillations as a function of the bias across the junction decays beyond 15 mV. The solid curve is a guide to the eye. (c) A schematic representation of carriers outside of the pseudogap of Gr$_\Delta$ entering Gr' when V$_{S-Gr}\gg \Delta_i$ across the junction.}
    \label{fig:Vd dependence}
\end{figure}

The typical sheet resistance ($R_{Gr}$) of graphene in our devices in response to the sweep up and sweep down (dual sweep) of gate voltage ($V_{bg}$) is shown in FIG.\ref{fig:Oscillations}-a. The resistance maxima for the Dirac point ($V_{DP}$) are often found to be between 0 V and +5 V, indicative of slight hole (p) doping from the substrate. $R_{Gr}$ is hysteresis-free at all temperatures, down to 4.2 K.
The four-terminal conductance of the S-Gr junction of device-A in FIG.\ref{fig:Oscillations}-b is measured as we sweep the gate which acts on the non-superconducting regions, Gr and Gr'. The gate voltage sets the Fermi level ($E_f$) in graphene as $E_f=sign(V_{bg}-V_{DP})\hbar v_f \sqrt{\pi C_{bg} |V_{bg}-V_{DP}|/e}$, where $\hbar$ is the reduced Planck constant, $v_f \sim1\times10^6$ m/s is the Fermi velocity in graphene , and $C_{bg}=38.37$ nF/cm$^2$ is the gate capacitance for 90 nm SiO$_2$. Note that the change in carrier density we can induce in BSCCO using our SiO$_2$/Si gate is not sufficient to affect its properties, unlike what is achievable by ionic gating \cite{Sterpetti2017,Liao2018}. In FIG.\ref{fig:Oscillations}-b, $E_f=0$ for $V_{DP}=+2.2$ V of Gr from our measurement of $R_{Gr}$. The extent of p-doping in Gr' is $E_f'=-180$ meV, which can be inferred from the second conductance minima in gate voltage at +12 V when the bias ($V_{S-Gr}$) is more than 25 meV in FIG.\ref{fig:Vd dependence}-a. The importance of the bias dependence is explained below. The presence of this p-doped Gr' region renders the two interfaces that are responsible for the resonant interaction between KT and AR in the junction, on which we will elaborate next. 

\begin{figure}[t!]
    \centering
    \includegraphics[width=\columnwidth]{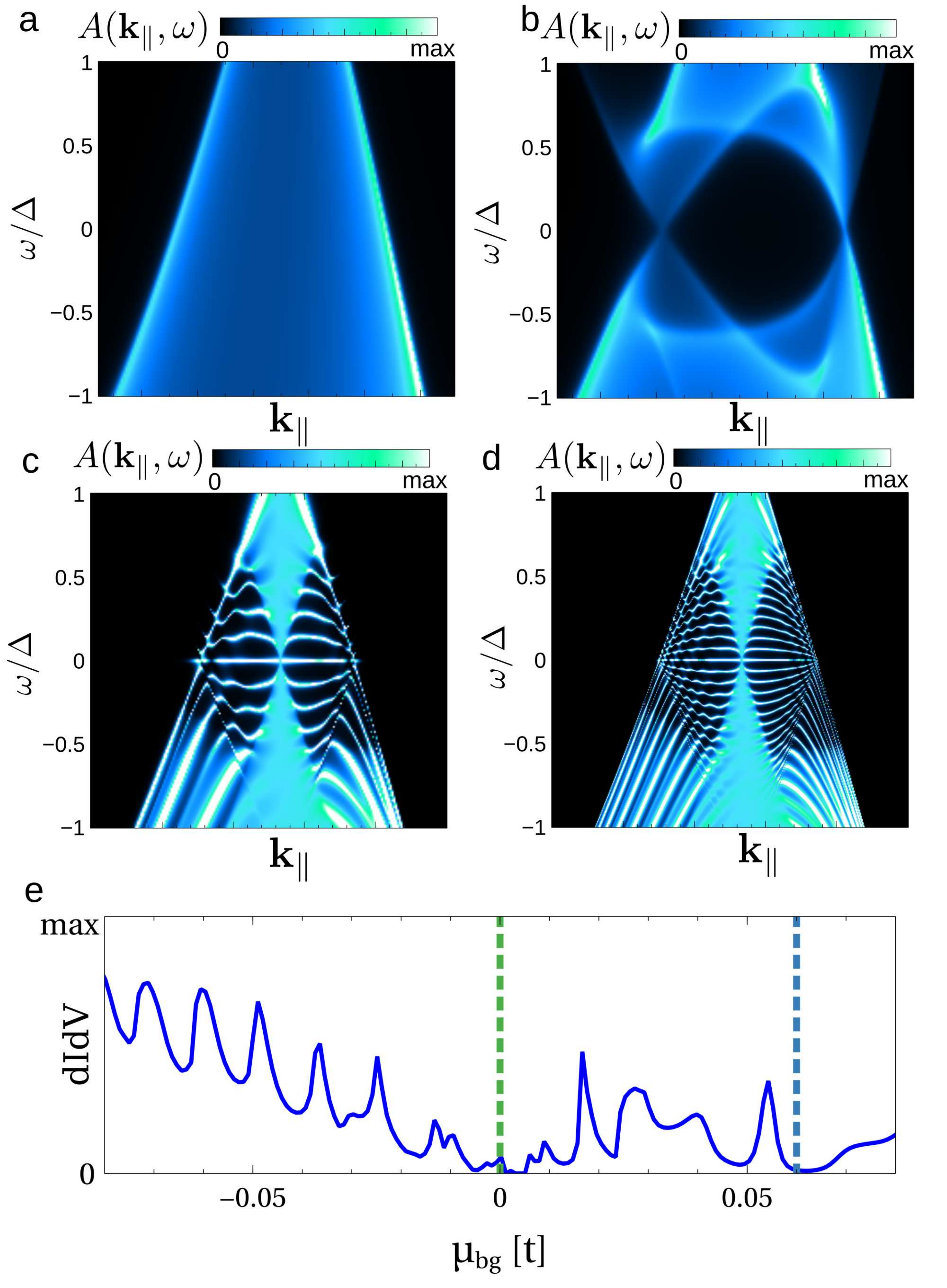}
    \caption{(a) Bulk spectral function for doped graphene, and doped graphene in the presence proximity induced d-wave superconductivity (b). Panel (c,d) shows the spectral function for the (c) unbiased and (d) biased gating conditions in the Gr' region. Panel (e) shows the computed dI/dV at $V_{S-Gr} = 0.07\Delta$, with the locations of the conductance minima of Dirac points marked by green ($\mu_{bg}=0$) and blue ($\mu_{bg}=0.06t$) dashed lines for the Gr and Gr' regions respectively. 
    }
    \label{fig:theory}
\end{figure}

The gate acts on Gr' and Gr regions simultaneously. When the gate is $E_f<0$, Gr'-Gr regions are in a p$^{+}$-p configuration. And when $0<E_f<-E_f'$ meV, Gr'-Gr regions are in a p-n configuration. The bottom panel of FIG.\ref{fig:Oscillations}-c shows the doping in the Gr$_\Delta$-Gr'-Gr regions relative to the Fermi-level (gray line), when Gr'-Gr regions are in a p-n configuration set by the gate. A nodal gap in Gr$_\Delta$ region comes from the proximity to the d-wave order parameter of BSCCO and is further supported by the calculated spectral function in FIG.\ref{fig:theory}-b. In FIG.\ref{fig:Oscillations}-b, the conductance asymmetry manifests as a 30\% decrease in the average conductance for the p-n configuration compared to the p-p configuration. This is due to the reduction in the transmission of carriers that are incident at oblique angles on the p-n junction \cite{Cheianov2006,Sutar2012}, owing to the chiral nature of carriers in graphene. The observation of the conductance asymmetry shown in FIG.\ref{fig:Oscillations}-b, and verified in FIG.\ref{fig:theory}-e, is one of the central results of this work. 

We note that the conductance asymmetry due to KT in this work is higher than that in a graphene p-n junction \cite{Sutar2012} due to the interactions with AR and the complicated angle dependence of each of these scattering mechanisms. All of these interactions are captured in the calculated conductance and spectral functions below. 

The other remarkable feature of FIG.\ref{fig:Oscillations}-b is the conductance oscillations which exist in both p$^+$-p and p-n doping configurations. The top panel of FIG.\ref{fig:Oscillations}-c shows the classical trajectories of carriers to visualize the number of round trips in Gr' region that lead to the resonant condition. Consider the p-n configuration in Gr'-Gr and a hole (dashed) entering Gr' by KT (top \textit{1}). It impinges on the Gr$_\Delta$-Gr' interface, and reflects as an electron (solid) by AR (\textit{2}). The electron (\textit{2}) makes a round trip (\textit{3}) by partial reflection from the Gr'-Gr interface. The returning electron (\textit{3}) travels to the Gr$_\Delta$ region and reflects as a hole (\textit{4}). A destructive resonant interference condition occurs every two round trips among carriers of the same charge, traversing in opposite directions in the Gr' region. The smaller width of the arrows in the Gr region indicates reduced transmission. The described resonant process is unchanged for p$^+$-p, except with higher transmission at the Gr'-Gr interface. The resonant mechanism is analogous to Fabry-P{\'{e}}rot (FP) resonances, but requires two full round trips ($4\times d$) as opposed to one round trip. The momentum-resolved spectral function in FIG.\ref{fig:theory}-c of the Gr' region in the Gr$_\Delta$-Gr'-Gr junction corroborates the formation of resonant modes. We note that tunneling into a nodal d-wave superconductor should lead to weaker AR compared to a system with an s-wave superconductor\cite{Linder2008,Bhattacharjee2006}. However, the significantly shorter coherence length of BSCCO compared to that of a conventional s-wave superconductor allows the formation of the three distinct regions needed to observe the resonances we report. On the other hand, quasi-periodic oscillations were observed in a Van der Waals NbSe$_2$/graphene junction \cite{Sahu2016}, whose origin was reported to be unclear.

The position of the troughs in FIG.\ref{fig:Oscillations}-b are highlighted in red. $E_{tr-tr}\sim17.3$ meV is the average period of oscillations. The small change in period between p$^+$-p and p-n doping configurations can be attributed to the change in $d$ with gating. We also note that resonances are less pronounced at low doping. The energy broadening $\delta\epsilon_f=$ 50 meV around the Dirac point from charge puddles in the SiO$_2$ \cite{Dean2010} washes out the oscillations. We use the oscillation period to quantify the length $d$ of the Gr' region using the equation below. 
\begin{equation}
d = \frac{hv_f}{4E_{tr-tr}}
\label{eq:d}
\end{equation}
From EQ-(\ref{eq:d}), the length of Gr' is $d\sim59$ nm. The values for $E_{tr-tr}$ and $d$ for all the devices in Table-\ref{tab:table1} are similar, regardless of other junction parameters. The mean free path $\lambda_{mfp}\sim80$ nm which we estimate, using $\sigma=(2e^2/h)k_f\lambda_{mfp}$, is slightly longer than the length $d$ of the Gr’ region which we estimate and is consistent with the phase coherent transport which we observe within the Gr’ region.

The contribution of AR to the resonance is evident by the decay in the oscillation amplitude when the bias is larger than the induced gap, as seen FIG.\ref{fig:Vd dependence}-a,b. FIG.\ref{fig:Vd dependence}-a is a plot of the junction conductance as the bias $V_{S-Gr}$ across the junction is increased. In FIG.\ref{fig:Vd dependence}-a, the increasing Fermi window allows the injection of carriers from outside of the induced superconducting gap in graphene, $\Delta_{i}$, (see FIG.\ref{fig:Vd dependence}-c) which suppresses the resonances. FIG. \ref{fig:Vd dependence}-b shows that the normalized oscillation amplitude decays when $V_{S-Gr}>15$ mV and implies $\Delta_i\sim15$ meV for device-A. We observe the same behavior in the other devices and inferred $\Delta_i$ from bias dependence ranges between 10-15 meV. This variation expected due to the sensitivity of the proximity-effect due to fabrication non-uniformities. Comparable magnitude of $\Delta_{i}$ has been seen in BSCCO heterojunctions with other 2D materials \cite{Wu2019,Zareapour2017}. The resonances are robust with temperature as shown in the supplemental information. This observation is consistent with the temperature dependence of $\Delta_{i}$ reported by others \cite{Wu2019}.

\begin{table}[b]
\begin{ruledtabular}
\caption{\label{tab:table1}
Properties of the junctions examined in this work: T$_C$ is the transition temperature of BSCCO, W the junction width, R*W the junction resistance, $E_{tr-tr}$ the period of oscillations, and $d$ the width of the Gr' region.
}
\begin{tabular}{cccccc}
\textrm{Device}&
\textrm{$T_c$}&
\textrm{W}&
\textrm{$R\times W$}&
% \textrm{U$_0$}&
\textrm{$E_{tr-tr}$}&
\textrm{$d$}\\

 & (K) & ($\mu m$) & (k$\Omega$-$\mu$m) & (meV) & (nm)\\
\colrule
A & 85 & 28.2 & 628.24 & 17.3 & 59\\ % BSCCO-Gr X stacks, Dev3
B & 84 & 26.8 & 639.23 & 16.1 & 64\\ % BSCCO-Gr Cracked stacks_3, Dev5
C & 70 & 4.8 & 29.85 & 15.9 & 65\\ % BSCCO-Gr Cracked stacks_6, Dev5
D & 85 & 5.8 & 92.85 & 19.4 & 53\\ % BSCCO-Gr Cracked stacks_3, Dev3
E & 85 & 4.0 & 36.00 & 17.3 & 60\\ % BSCCO-Gr Cracked Stacks_3, Dev7(Gr)
% F & 82 & 11.6 & 0.19 & 21.7 & 47\\ % BSCCO-Gr Cracked Stacks_6, Dev6 
\end{tabular}

\end{ruledtabular}
\end{table}

To capture the electronic properties of the junction, we take an effective low-energy model of the form
$
    \mathcal{H} =  \mathcal{H}_{\text{kin}} + \mathcal{H}_{\text{SC}} +    \mathcal{H}_{\mu}
$
with 
$
     \mathcal{H}_{\text{kin}} = 
     t \sum_{\langle ij \rangle,s} 
     c^\dagger_{i,s} c_{j,s} +h.c.
$
capturing the kinetic energy of electrons, 
where $c^\dagger_{i,s}$ denotes the creation operator in site $i$ with spin $s$, and
$\langle \rangle$ denotes sum over first neighbors. The term
$
     \mathcal{H}_{\text{SC}} = 
     \sum_{\langle ij \rangle \in \text{Gr}_\Delta} \Delta_{ij} c_{i,\uparrow} c_{j,\downarrow} + \text{h.c.}
$
accounts for the superconducting proximity effect,
where $\Delta_{ij}$ is defined such that its Fourier transform
leads to a $d_{x^2-y^2}$ order parameter in reciprocal space.
Finally, the term
$
     \mathcal{H}_{\mu} = 
     \mu_\Delta
     \sum_{i \in \text{Gr}_\Delta,s} 
     c^\dagger_{i,s} c_{i,s} 
     +
     \mu'_{\text{bg}}
     \sum_{i \in \text{Gr'},s} 
     c^\dagger_{i,s} c_{i,s} 
     +
          \mu_{\text{bg}}
     \sum_{i \in \text{Gr},s} 
     c^\dagger_{i,s} c_{i,s} 
$
controls the local chemical potential in the different regions, to account for the experimental conditions. 
For concreteness we take that the screening in the Gr and Gr' regions is similar
yielding $\mu_{\text{bg}}=0.06t + \mu'_{\text{bg}}$, 
which would depend on the experimentally applied bias $V_{\text{bg}}$,
and the shift corresponds to the difference in doping.
$\mu_\Delta$ arises due to charge transfer from BSCCO,
and we take it independent of the bias due to screening.
We define a junction that is infinite in the $y$-direction and a finite Gr' is defined between the two semi-infinite Gr$_\Delta$ and Gr regions, as depicted in FIG.\ref{fig:Oscillations}-c. Due to the translation symmetry in $y$, the spectral function of the junction can be labeled according to the parallel momenta $\mathbf{k}_\parallel$ to the interface. The previous Hamiltonian is solved with the Bogoliubov-de Gennes formalism, and we compute the electron spectral function at Gr' defined as $A(\mathbf{k}_\parallel,\omega) = 
-\frac{1}{\pi} \text{Tr}_{\text{Gr'}} \text{Im} \left ( P_e \mathcal{G}(\mathbf{k}_\parallel,\omega)\right )$, where $P_e$ is the projection operator in the electron Nambu sector and $\mathcal{G}(\mathbf{k}_\parallel,\omega)$ is the full Nambu Green's function of the system. $A(\mathbf{k}_\parallel,\omega)$ is used to directly extract the electronic excitation emerging in the junction and is computed with the Green's function re-normalization algorithm \cite{Sancho1985}.

First, we show the spectral function versus the transverse momenta in doped graphene in the absence (FIG.\ref{fig:theory}-a) and presence (FIG.\ref{fig:theory}-b) of a d-wave superconducting proximity effect. It is observed that the proximity to a d-wave superconductor opens up a momentum-dependent electronic structure in graphene (FIG. \ref{fig:theory}-b). Panel (FIG.\ref{fig:theory}-c,d) shows the momentum-resolved spectral function at the center of the Gr' region, featuring the emergence of resonant modes. The increase in oscillatory single-particle states of FIG.\ref{fig:theory}-d shows the sensitivity of the resonances to gating. The narrowing of the conductance oscillation period at higher $E_f$ seen in FIG\ref{fig:Oscillations}-b is due to the increase in the number of resonant modes.

To show the impact of those resonances in transport, we compute the conductance of the entire junction $G(\omega)$ using the non-perturbative $\mathcal{S}$-matrix formalism \cite{Datta1995,Blonder1982,PhysRevB.23.6851,Lado}. The resultant conductance as we vary the Fermi energy in the Gr'-Gr regions is shown in FIG.\ref{fig:theory}-e for low excitation energy. The minimum conductance at $E = 0$, marked by the green dashed line corresponds to the Fermi level of the Gr region. The conductance asymmetry between the p$^+$-p configuration ($E>0$) versus p-n configuration ($E>0$) of the Gr'-Gr regions is clearly evident, as also seen in the experiment. 

To summarize, we show the appearance of resonances in the transport properties of BSCCO/graphene junctions due to two different scattering mechanisms at two interfaces that form due to charge transfer and superconducting proximity. The first interface gives rise to Andreev reflection and occurs between a region of superconducting graphene and a region of graphene that is doped by charge transfer from BSCCO. The second interface occurs between the doped region and the remainder of native graphene and gives rise to Klein tunneling, which manifests in a conductance asymmetry. The resonant interference occurs for every two round trips of carriers within the doped region. Our results are supported by the computed momentum-resolved spectral functions and conductance. Our results have shown unique transport phenomena in high-temperature superconductor/graphene junctions that will lead to better understanding of future devices such as Josephson junctions. 

This work was supported by funding from the U.S. Naval Research Laboratory grant N00173-19-1-G0008. JLL acknowledges the computational resources provided by the Aalto Science-IT project, and the financial support from the Academy of Finland Projects No. 331342 and No. 336243. The work at Brookhaven National Laboratory was supported by the U.S. Department of Energy (DOE) the Office of Basic Energy Sciences, Materials Sciences, and Engineering Division under Contract No. DESC0012704.

\bibliography{biblio}

\begin{thebibliography}{39}%
\makeatletter
\providecommand \@ifxundefined [1]{%
 \@ifx{#1\undefined}
}%
\providecommand \@ifnum [1]{%
 \ifnum #1\expandafter \@firstoftwo
 \else \expandafter \@secondoftwo
 \fi
}%
\providecommand \@ifx [1]{%
 \ifx #1\expandafter \@firstoftwo
 \else \expandafter \@secondoftwo
 \fi
}%
\providecommand \natexlab [1]{#1}%
\providecommand \enquote  [1]{``#1''}%
\providecommand \bibnamefont  [1]{#1}%
\providecommand \bibfnamefont [1]{#1}%
\providecommand \citenamefont [1]{#1}%
\providecommand \href@noop [0]{\@secondoftwo}%
\providecommand \href [0]{\begingroup \@sanitize@url \@href}%
\providecommand \@href[1]{\@@startlink{#1}\@@href}%
\providecommand \@@href[1]{\endgroup#1\@@endlink}%
\providecommand \@sanitize@url [0]{\catcode `\\12\catcode `\$12\catcode
  `\&12\catcode `\#12\catcode `\^12\catcode `\_12\catcode `\%12\relax}%
\providecommand \@@startlink[1]{}%
\providecommand \@@endlink[0]{}%
\providecommand \url  [0]{\begingroup\@sanitize@url \@url }%
\providecommand \@url [1]{\endgroup\@href {#1}{\urlprefix }}%
\providecommand \urlprefix  [0]{URL }%
\providecommand \Eprint [0]{\href }%
\providecommand \doibase [0]{https://doi.org/}%
\providecommand \selectlanguage [0]{\@gobble}%
\providecommand \bibinfo  [0]{\@secondoftwo}%
\providecommand \bibfield  [0]{\@secondoftwo}%
\providecommand \translation [1]{[#1]}%
\providecommand \BibitemOpen [0]{}%
\providecommand \bibitemStop [0]{}%
\providecommand \bibitemNoStop [0]{.\EOS\space}%
\providecommand \EOS [0]{\spacefactor3000\relax}%
\providecommand \BibitemShut  [1]{\csname bibitem#1\endcsname}%
\let\auto@bib@innerbib\@empty
%</preamble>
\bibitem [{\citenamefont {Rowell}\ and\ \citenamefont
  {McMillan}(1966)}]{Rowell1966}%
  \BibitemOpen
  \bibfield  {author} {\bibinfo {author} {\bibfnamefont {J.~M.}\ \bibnamefont
  {Rowell}}\ and\ \bibinfo {author} {\bibfnamefont {W.~L.}\ \bibnamefont
  {McMillan}},\ }\bibfield  {title} {\bibinfo {title} {{Electron interference
  in a normal metal induced by superconducting contacts}},\ }\href
  {https://doi.org/10.1103/PhysRevLett.16.453} {\bibfield  {journal} {\bibinfo
  {journal} {Physical Review Letters}\ }\textbf {\bibinfo {volume} {16}},\
  \bibinfo {pages} {453} (\bibinfo {year} {1966})}\BibitemShut {NoStop}%
\bibitem [{\citenamefont {L{\"{o}}fwander}\ \emph {et~al.}(2001)\citenamefont
  {L{\"{o}}fwander}, \citenamefont {Shumeiko},\ and\ \citenamefont
  {Wendin}}]{Lofwander2001}%
  \BibitemOpen
  \bibfield  {author} {\bibinfo {author} {\bibfnamefont {T.}~\bibnamefont
  {L{\"{o}}fwander}}, \bibinfo {author} {\bibfnamefont {V.~S.}\ \bibnamefont
  {Shumeiko}},\ and\ \bibinfo {author} {\bibfnamefont {G.}~\bibnamefont
  {Wendin}},\ }\bibfield  {title} {\bibinfo {title} {{Andreev bound states in
  high-T$_c$ superconducting junctions}},\ }\href
  {https://doi.org/10.1088/0953-2048/14/5/201} {\bibfield  {journal} {\bibinfo
  {journal} {Superconductor Science and Technology}\ }\textbf {\bibinfo
  {volume} {14}},\ \bibinfo {pages} {R53} (\bibinfo {year} {2001})},\ \Eprint
  {https://arxiv.org/abs/0102276} {arXiv:0102276 [cond-mat]} \BibitemShut
  {NoStop}%
\bibitem [{\citenamefont {Dirks}\ \emph {et~al.}(2011)\citenamefont {Dirks},
  \citenamefont {Hughes}, \citenamefont {Lal}, \citenamefont {Uchoa},
  \citenamefont {Chen}, \citenamefont {Chialvo}, \citenamefont {Goldbart},\
  and\ \citenamefont {Mason}}]{Dirks2011}%
  \BibitemOpen
  \bibfield  {author} {\bibinfo {author} {\bibfnamefont {T.}~\bibnamefont
  {Dirks}}, \bibinfo {author} {\bibfnamefont {T.~L.}\ \bibnamefont {Hughes}},
  \bibinfo {author} {\bibfnamefont {S.}~\bibnamefont {Lal}}, \bibinfo {author}
  {\bibfnamefont {B.}~\bibnamefont {Uchoa}}, \bibinfo {author} {\bibfnamefont
  {Y.~F.}\ \bibnamefont {Chen}}, \bibinfo {author} {\bibfnamefont
  {C.}~\bibnamefont {Chialvo}}, \bibinfo {author} {\bibfnamefont {P.~M.}\
  \bibnamefont {Goldbart}},\ and\ \bibinfo {author} {\bibfnamefont
  {N.}~\bibnamefont {Mason}},\ }\bibfield  {title} {\bibinfo {title}
  {{Transport through Andreev bound states in a graphene quantum dot}},\ }\href
  {https://doi.org/10.1038/nphys1911} {\bibfield  {journal} {\bibinfo
  {journal} {Nature Physics}\ }\textbf {\bibinfo {volume} {7}},\ \bibinfo
  {pages} {386} (\bibinfo {year} {2011})},\ \Eprint
  {https://arxiv.org/abs/1005.2749} {arXiv:1005.2749} \BibitemShut {NoStop}%
\bibitem [{\citenamefont {Novoselov}\ \emph {et~al.}(2005)\citenamefont
  {Novoselov}, \citenamefont {Geim}, \citenamefont {Morozov}, \citenamefont
  {Jiang}, \citenamefont {Katsnelson}, \citenamefont {Grigorieva},
  \citenamefont {Dubonos},\ and\ \citenamefont {Firsov}}]{Novoselov2005}%
  \BibitemOpen
  \bibfield  {author} {\bibinfo {author} {\bibfnamefont {K.~S.}\ \bibnamefont
  {Novoselov}}, \bibinfo {author} {\bibfnamefont {A.~K.}\ \bibnamefont {Geim}},
  \bibinfo {author} {\bibfnamefont {S.~V.}\ \bibnamefont {Morozov}}, \bibinfo
  {author} {\bibfnamefont {D.}~\bibnamefont {Jiang}}, \bibinfo {author}
  {\bibfnamefont {M.~I.}\ \bibnamefont {Katsnelson}}, \bibinfo {author}
  {\bibfnamefont {I.~V.}\ \bibnamefont {Grigorieva}}, \bibinfo {author}
  {\bibfnamefont {S.~V.}\ \bibnamefont {Dubonos}},\ and\ \bibinfo {author}
  {\bibfnamefont {A.~A.}\ \bibnamefont {Firsov}},\ }\bibfield  {title}
  {\bibinfo {title} {{Two-dimensional gas of massless Dirac fermions in
  graphene}},\ }\bibfield  {journal} {\bibinfo  {journal} {Nature}\ }\href
  {https://doi.org/10.1038/nature04233} {10.1038/nature04233} (\bibinfo {year}
  {2005}),\ \Eprint {https://arxiv.org/abs/0509330} {arXiv:0509330 [cond-mat]}
  \BibitemShut {NoStop}%
\bibitem [{\citenamefont {Hamidian}\ \emph {et~al.}(2016)\citenamefont
  {Hamidian}, \citenamefont {Edkins}, \citenamefont {Joo}, \citenamefont
  {Kostin}, \citenamefont {Eisaki}, \citenamefont {Uchida}, \citenamefont
  {Lawler}, \citenamefont {Kim}, \citenamefont {Mackenzie}, \citenamefont
  {Fujita}, \citenamefont {Lee},\ and\ \citenamefont {Davis}}]{Hamidian2016}%
  \BibitemOpen
  \bibfield  {author} {\bibinfo {author} {\bibfnamefont {M.~H.}\ \bibnamefont
  {Hamidian}}, \bibinfo {author} {\bibfnamefont {S.~D.}\ \bibnamefont
  {Edkins}}, \bibinfo {author} {\bibfnamefont {S.~H.}\ \bibnamefont {Joo}},
  \bibinfo {author} {\bibfnamefont {A.}~\bibnamefont {Kostin}}, \bibinfo
  {author} {\bibfnamefont {H.}~\bibnamefont {Eisaki}}, \bibinfo {author}
  {\bibfnamefont {S.}~\bibnamefont {Uchida}}, \bibinfo {author} {\bibfnamefont
  {M.~J.}\ \bibnamefont {Lawler}}, \bibinfo {author} {\bibfnamefont {E.~A.}\
  \bibnamefont {Kim}}, \bibinfo {author} {\bibfnamefont {A.~P.}\ \bibnamefont
  {Mackenzie}}, \bibinfo {author} {\bibfnamefont {K.}~\bibnamefont {Fujita}},
  \bibinfo {author} {\bibfnamefont {J.}~\bibnamefont {Lee}},\ and\ \bibinfo
  {author} {\bibfnamefont {J.~C.}\ \bibnamefont {Davis}},\ }\bibfield  {title}
  {\bibinfo {title} {{Detection of a Cooper-pair density wave in
  Bi$_2$Sr$_2$CaCu$_2$O$_{8+\delta}$}},\ }\href
  {https://doi.org/10.1038/nature17411} {\bibfield  {journal} {\bibinfo
  {journal} {Nature}\ }\textbf {\bibinfo {volume} {532}},\ \bibinfo {pages}
  {343} (\bibinfo {year} {2016})},\ \Eprint {https://arxiv.org/abs/1511.08124}
  {arXiv:1511.08124} \BibitemShut {NoStop}%
\bibitem [{\citenamefont {Yu}\ \emph {et~al.}(2019)\citenamefont {Yu},
  \citenamefont {Ma}, \citenamefont {Cai}, \citenamefont {Zhong}, \citenamefont
  {Ye}, \citenamefont {Shen}, \citenamefont {Gu}, \citenamefont {Chen},\ and\
  \citenamefont {Zhang}}]{Yu2019}%
  \BibitemOpen
  \bibfield  {author} {\bibinfo {author} {\bibfnamefont {Y.}~\bibnamefont
  {Yu}}, \bibinfo {author} {\bibfnamefont {L.}~\bibnamefont {Ma}}, \bibinfo
  {author} {\bibfnamefont {P.}~\bibnamefont {Cai}}, \bibinfo {author}
  {\bibfnamefont {R.}~\bibnamefont {Zhong}}, \bibinfo {author} {\bibfnamefont
  {C.}~\bibnamefont {Ye}}, \bibinfo {author} {\bibfnamefont {J.}~\bibnamefont
  {Shen}}, \bibinfo {author} {\bibfnamefont {G.~D.}\ \bibnamefont {Gu}},
  \bibinfo {author} {\bibfnamefont {X.~H.}\ \bibnamefont {Chen}},\ and\
  \bibinfo {author} {\bibfnamefont {Y.}~\bibnamefont {Zhang}},\ }\bibfield
  {title} {\bibinfo {title} {{High-temperature superconductivity in monolayer
  Bi$_2$Sr$_2$CaCu$_2$O$_{8+\delta}$}},\ }\href
  {https://doi.org/10.1038/s41586-019-1718-x} {\bibfield  {journal} {\bibinfo
  {journal} {Nature}\ }\textbf {\bibinfo {volume} {575}},\ \bibinfo {pages}
  {156} (\bibinfo {year} {2019})}\BibitemShut {NoStop}%
\bibitem [{\citenamefont {Liao}\ \emph {et~al.}(2018)\citenamefont {Liao},
  \citenamefont {Zhu}, \citenamefont {Zhang}, \citenamefont {Zhong},
  \citenamefont {Schneeloch}, \citenamefont {Gu}, \citenamefont {Jiang},
  \citenamefont {Zhang}, \citenamefont {Ma},\ and\ \citenamefont
  {Xue}}]{Liao2018}%
  \BibitemOpen
  \bibfield  {author} {\bibinfo {author} {\bibfnamefont {M.}~\bibnamefont
  {Liao}}, \bibinfo {author} {\bibfnamefont {Y.}~\bibnamefont {Zhu}}, \bibinfo
  {author} {\bibfnamefont {J.}~\bibnamefont {Zhang}}, \bibinfo {author}
  {\bibfnamefont {R.}~\bibnamefont {Zhong}}, \bibinfo {author} {\bibfnamefont
  {J.}~\bibnamefont {Schneeloch}}, \bibinfo {author} {\bibfnamefont
  {G.}~\bibnamefont {Gu}}, \bibinfo {author} {\bibfnamefont {K.}~\bibnamefont
  {Jiang}}, \bibinfo {author} {\bibfnamefont {D.}~\bibnamefont {Zhang}},
  \bibinfo {author} {\bibfnamefont {X.}~\bibnamefont {Ma}},\ and\ \bibinfo
  {author} {\bibfnamefont {Q.~K.}\ \bibnamefont {Xue}},\ }\bibfield  {title}
  {\bibinfo {title} {{Superconductor-Insulator Transitions in Exfoliated
  Bi$_2$Sr$_2$CaCu$_2$O$_{8+\delta}$ Flakes}},\ }\href
  {https://doi.org/10.1021/acs.nanolett.8b02183} {\bibfield  {journal}
  {\bibinfo  {journal} {Nano Letters}\ }\textbf {\bibinfo {volume} {18}},\
  \bibinfo {pages} {5660} (\bibinfo {year} {2018})}\BibitemShut {NoStop}%
\bibitem [{\citenamefont {Wu}\ \emph {et~al.}(2019)\citenamefont {Wu},
  \citenamefont {Xiao}, \citenamefont {Li}, \citenamefont {Li}, \citenamefont
  {Li}, \citenamefont {Mu}, \citenamefont {Jiang}, \citenamefont {Hu},\ and\
  \citenamefont {Xie}}]{Wu2019}%
  \BibitemOpen
  \bibfield  {author} {\bibinfo {author} {\bibfnamefont {Y.}~\bibnamefont
  {Wu}}, \bibinfo {author} {\bibfnamefont {H.}~\bibnamefont {Xiao}}, \bibinfo
  {author} {\bibfnamefont {Q.}~\bibnamefont {Li}}, \bibinfo {author}
  {\bibfnamefont {X.}~\bibnamefont {Li}}, \bibinfo {author} {\bibfnamefont
  {Z.}~\bibnamefont {Li}}, \bibinfo {author} {\bibfnamefont {G.}~\bibnamefont
  {Mu}}, \bibinfo {author} {\bibfnamefont {D.}~\bibnamefont {Jiang}}, \bibinfo
  {author} {\bibfnamefont {T.}~\bibnamefont {Hu}},\ and\ \bibinfo {author}
  {\bibfnamefont {X.~M.}\ \bibnamefont {Xie}},\ }\bibfield  {title} {\bibinfo
  {title} {{The transport properties in graphene/single-unit-cell cuprates van
  der Waals heterostructure}},\ }\bibfield  {journal} {\bibinfo  {journal}
  {Superconductor Science and Technology}\ }\textbf {\bibinfo {volume} {32}},\
  \href {https://doi.org/10.1088/1361-6668/ab2097} {10.1088/1361-6668/ab2097}
  (\bibinfo {year} {2019})\BibitemShut {NoStop}%
\bibitem [{\citenamefont {Zareapour}\ \emph {et~al.}(2012)\citenamefont
  {Zareapour}, \citenamefont {Hayat}, \citenamefont {Zhao}, \citenamefont
  {Kreshchuk}, \citenamefont {Jain}, \citenamefont {Kwok}, \citenamefont {Lee},
  \citenamefont {Cheong}, \citenamefont {Xu}, \citenamefont {Yang},
  \citenamefont {Gu}, \citenamefont {Jia}, \citenamefont {Cava},\ and\
  \citenamefont {Burch}}]{Zareapour2012}%
  \BibitemOpen
  \bibfield  {author} {\bibinfo {author} {\bibfnamefont {P.}~\bibnamefont
  {Zareapour}}, \bibinfo {author} {\bibfnamefont {A.}~\bibnamefont {Hayat}},
  \bibinfo {author} {\bibfnamefont {S.~Y.~F.}\ \bibnamefont {Zhao}}, \bibinfo
  {author} {\bibfnamefont {M.}~\bibnamefont {Kreshchuk}}, \bibinfo {author}
  {\bibfnamefont {A.}~\bibnamefont {Jain}}, \bibinfo {author} {\bibfnamefont
  {D.~C.}\ \bibnamefont {Kwok}}, \bibinfo {author} {\bibfnamefont
  {N.}~\bibnamefont {Lee}}, \bibinfo {author} {\bibfnamefont {S.~W.}\
  \bibnamefont {Cheong}}, \bibinfo {author} {\bibfnamefont {Z.}~\bibnamefont
  {Xu}}, \bibinfo {author} {\bibfnamefont {A.}~\bibnamefont {Yang}}, \bibinfo
  {author} {\bibfnamefont {G.~D.}\ \bibnamefont {Gu}}, \bibinfo {author}
  {\bibfnamefont {S.}~\bibnamefont {Jia}}, \bibinfo {author} {\bibfnamefont
  {R.~J.}\ \bibnamefont {Cava}},\ and\ \bibinfo {author} {\bibfnamefont
  {K.~S.}\ \bibnamefont {Burch}},\ }\bibfield  {title} {\bibinfo {title}
  {{Proximity-induced high-temperature superconductivity in the topological
  insulators Bi$_2$Se$_3$ and Bi$_2$Te$_3$}},\ }\bibfield  {journal} {\bibinfo
  {journal} {Nature Communications}\ }\textbf {\bibinfo {volume} {3}},\ \href
  {https://doi.org/10.1038/ncomms2042} {10.1038/ncomms2042} (\bibinfo {year}
  {2012})\BibitemShut {NoStop}%
\bibitem [{\citenamefont {Zareapour}\ \emph {et~al.}(2017)\citenamefont
  {Zareapour}, \citenamefont {Hayat}, \citenamefont {Zhao}, \citenamefont
  {Kreshchuk}, \citenamefont {Xu}, \citenamefont {Liu}, \citenamefont {Gu},
  \citenamefont {Jia}, \citenamefont {Cava}, \citenamefont {Yang},
  \citenamefont {Ran},\ and\ \citenamefont {Burch}}]{Zareapour2017}%
  \BibitemOpen
  \bibfield  {author} {\bibinfo {author} {\bibfnamefont {P.}~\bibnamefont
  {Zareapour}}, \bibinfo {author} {\bibfnamefont {A.}~\bibnamefont {Hayat}},
  \bibinfo {author} {\bibfnamefont {S.~Y.~F.}\ \bibnamefont {Zhao}}, \bibinfo
  {author} {\bibfnamefont {M.}~\bibnamefont {Kreshchuk}}, \bibinfo {author}
  {\bibfnamefont {Z.}~\bibnamefont {Xu}}, \bibinfo {author} {\bibfnamefont
  {T.~S.}\ \bibnamefont {Liu}}, \bibinfo {author} {\bibfnamefont {G.~D.}\
  \bibnamefont {Gu}}, \bibinfo {author} {\bibfnamefont {S.}~\bibnamefont
  {Jia}}, \bibinfo {author} {\bibfnamefont {R.~J.}\ \bibnamefont {Cava}},
  \bibinfo {author} {\bibfnamefont {H.-Y.}\ \bibnamefont {Yang}}, \bibinfo
  {author} {\bibfnamefont {Y.}~\bibnamefont {Ran}},\ and\ \bibinfo {author}
  {\bibfnamefont {K.~S.}\ \bibnamefont {Burch}},\ }\bibfield  {title} {\bibinfo
  {title} {{Andreev reflection without Fermi surface alignment in high- T c van
  der Waals heterostructures}},\ }\href
  {https://doi.org/10.1088/1367-2630/aa6b8a} {\bibfield  {journal} {\bibinfo
  {journal} {New Journal of Physics}\ }\textbf {\bibinfo {volume} {19}},\
  \bibinfo {pages} {043026} (\bibinfo {year} {2017})}\BibitemShut {NoStop}%
\bibitem [{\citenamefont {{Di Bernardo}}\ \emph {et~al.}(2017)\citenamefont
  {{Di Bernardo}}, \citenamefont {Millo}, \citenamefont {Barbone},
  \citenamefont {Alpern}, \citenamefont {Kalcheim}, \citenamefont {Sassi},
  \citenamefont {Ott}, \citenamefont {{De Fazio}}, \citenamefont {Yoon},
  \citenamefont {Amado}, \citenamefont {Ferrari}, \citenamefont {Linder},\ and\
  \citenamefont {Robinson}}]{DiBernardo2017}%
  \BibitemOpen
  \bibfield  {author} {\bibinfo {author} {\bibfnamefont {A.}~\bibnamefont {{Di
  Bernardo}}}, \bibinfo {author} {\bibfnamefont {O.}~\bibnamefont {Millo}},
  \bibinfo {author} {\bibfnamefont {M.}~\bibnamefont {Barbone}}, \bibinfo
  {author} {\bibfnamefont {H.}~\bibnamefont {Alpern}}, \bibinfo {author}
  {\bibfnamefont {Y.}~\bibnamefont {Kalcheim}}, \bibinfo {author}
  {\bibfnamefont {U.}~\bibnamefont {Sassi}}, \bibinfo {author} {\bibfnamefont
  {A.~K.}\ \bibnamefont {Ott}}, \bibinfo {author} {\bibfnamefont
  {D.}~\bibnamefont {{De Fazio}}}, \bibinfo {author} {\bibfnamefont
  {D.}~\bibnamefont {Yoon}}, \bibinfo {author} {\bibfnamefont {M.}~\bibnamefont
  {Amado}}, \bibinfo {author} {\bibfnamefont {A.~C.}\ \bibnamefont {Ferrari}},
  \bibinfo {author} {\bibfnamefont {J.}~\bibnamefont {Linder}},\ and\ \bibinfo
  {author} {\bibfnamefont {J.~W.}\ \bibnamefont {Robinson}},\ }\bibfield
  {title} {\bibinfo {title} {{P-wave triggered superconductivity in
  single-layer graphene on an electron-doped oxide superconductor}},\
  }\bibfield  {journal} {\bibinfo  {journal} {Nature Communications}\ }\textbf
  {\bibinfo {volume} {8}},\ \href {https://doi.org/10.1038/ncomms14024}
  {10.1038/ncomms14024} (\bibinfo {year} {2017})\BibitemShut {NoStop}%
\bibitem [{\citenamefont {Black-Schaffer}\ and\ \citenamefont
  {Honerkamp}(2014)}]{Black-Schaffer2014}%
  \BibitemOpen
  \bibfield  {author} {\bibinfo {author} {\bibfnamefont {A.~M.}\ \bibnamefont
  {Black-Schaffer}}\ and\ \bibinfo {author} {\bibfnamefont {C.}~\bibnamefont
  {Honerkamp}},\ }\bibfield  {title} {\bibinfo {title} {{Chiral d -wave
  superconductivity in doped graphene}},\ }\href
  {https://doi.org/10.1088/0953-8984/26/42/423201} {\bibfield  {journal}
  {\bibinfo  {journal} {Journal of Physics: Condensed Matter}\ }\textbf
  {\bibinfo {volume} {26}},\ \bibinfo {pages} {423201} (\bibinfo {year}
  {2014})},\ \Eprint {https://arxiv.org/abs/1406.0101} {arXiv:1406.0101}
  \BibitemShut {NoStop}%
\bibitem [{\citenamefont {Katsnelson}\ \emph {et~al.}(2006)\citenamefont
  {Katsnelson}, \citenamefont {Novoselov},\ and\ \citenamefont
  {Geim}}]{Katsnelson2006}%
  \BibitemOpen
  \bibfield  {author} {\bibinfo {author} {\bibfnamefont {M.~I.}\ \bibnamefont
  {Katsnelson}}, \bibinfo {author} {\bibfnamefont {K.~S.}\ \bibnamefont
  {Novoselov}},\ and\ \bibinfo {author} {\bibfnamefont {A.~K.}\ \bibnamefont
  {Geim}},\ }\bibfield  {title} {\bibinfo {title} {{Chiral tunnelling and the
  Klein paradox in graphene}},\ }\href {https://doi.org/10.1038/nphys384}
  {\bibfield  {journal} {\bibinfo  {journal} {Nature Physics}\ }\textbf
  {\bibinfo {volume} {2}},\ \bibinfo {pages} {620} (\bibinfo {year} {2006})},\
  \Eprint {https://arxiv.org/abs/0604323} {arXiv:0604323 [cond-mat]}
  \BibitemShut {NoStop}%
\bibitem [{\citenamefont {Cheianov}\ and\ \citenamefont
  {Fal'ko}(2006)}]{Cheianov2006}%
  \BibitemOpen
  \bibfield  {author} {\bibinfo {author} {\bibfnamefont {V.~V.}\ \bibnamefont
  {Cheianov}}\ and\ \bibinfo {author} {\bibfnamefont {V.~I.}\ \bibnamefont
  {Fal'ko}},\ }\bibfield  {title} {\bibinfo {title} {Selective transmission of
  dirac electrons and ballistic magnetoresistance of $n\text{\ensuremath{-}}p$
  junctions in graphene},\ }\href {https://doi.org/10.1103/PhysRevB.74.041403}
  {\bibfield  {journal} {\bibinfo  {journal} {Phys. Rev. B}\ }\textbf {\bibinfo
  {volume} {74}},\ \bibinfo {pages} {041403(R)} (\bibinfo {year}
  {2006})}\BibitemShut {NoStop}%
\bibitem [{\citenamefont {Reijnders}\ \emph {et~al.}(2013)\citenamefont
  {Reijnders}, \citenamefont {Tudorovskiy},\ and\ \citenamefont
  {Katsnelson}}]{Reijnders2013}%
  \BibitemOpen
  \bibfield  {author} {\bibinfo {author} {\bibfnamefont {K.}~\bibnamefont
  {Reijnders}}, \bibinfo {author} {\bibfnamefont {T.}~\bibnamefont
  {Tudorovskiy}},\ and\ \bibinfo {author} {\bibfnamefont {M.}~\bibnamefont
  {Katsnelson}},\ }\bibfield  {title} {\bibinfo {title} {{Semiclassical theory
  of potential scattering for massless Dirac fermions}},\ }\href
  {https://doi.org/10.1016/j.aop.2013.03.001} {\bibfield  {journal} {\bibinfo
  {journal} {Annals of Physics}\ }\textbf {\bibinfo {volume} {333}},\ \bibinfo
  {pages} {155} (\bibinfo {year} {2013})},\ \Eprint
  {https://arxiv.org/abs/1206.2869} {arXiv:1206.2869} \BibitemShut {NoStop}%
\bibitem [{\citenamefont {Beenakker}(2008)}]{Beenakker2008}%
  \BibitemOpen
  \bibfield  {author} {\bibinfo {author} {\bibfnamefont {C.~W.~J.}\
  \bibnamefont {Beenakker}},\ }\bibfield  {title} {\bibinfo {title}
  {Colloquium: Andreev reflection and klein tunneling in graphene},\ }\href
  {https://doi.org/10.1103/RevModPhys.80.1337} {\bibfield  {journal} {\bibinfo
  {journal} {Rev. Mod. Phys.}\ }\textbf {\bibinfo {volume} {80}},\ \bibinfo
  {pages} {1337} (\bibinfo {year} {2008})}\BibitemShut {NoStop}%
\bibitem [{\citenamefont {Burset}\ \emph {et~al.}(2009)\citenamefont {Burset},
  \citenamefont {Herrera},\ and\ \citenamefont {Levy~Yeyati}}]{Burset2009}%
  \BibitemOpen
  \bibfield  {author} {\bibinfo {author} {\bibfnamefont {P.}~\bibnamefont
  {Burset}}, \bibinfo {author} {\bibfnamefont {W.}~\bibnamefont {Herrera}},\
  and\ \bibinfo {author} {\bibfnamefont {A.}~\bibnamefont {Levy~Yeyati}},\
  }\bibfield  {title} {\bibinfo {title} {Proximity-induced interface bound
  states in superconductor-graphene junctions},\ }\href
  {https://doi.org/10.1103/PhysRevB.80.041402} {\bibfield  {journal} {\bibinfo
  {journal} {Phys. Rev. B}\ }\textbf {\bibinfo {volume} {80}},\ \bibinfo
  {pages} {041402(R)} (\bibinfo {year} {2009})}\BibitemShut {NoStop}%
\bibitem [{\citenamefont {Huard}\ \emph {et~al.}(2007)\citenamefont {Huard},
  \citenamefont {Sulpizio}, \citenamefont {Stander}, \citenamefont {Todd},
  \citenamefont {Yang},\ and\ \citenamefont {Goldhaber-Gordon}}]{Huard2007}%
  \BibitemOpen
  \bibfield  {author} {\bibinfo {author} {\bibfnamefont {B.}~\bibnamefont
  {Huard}}, \bibinfo {author} {\bibfnamefont {J.~A.}\ \bibnamefont {Sulpizio}},
  \bibinfo {author} {\bibfnamefont {N.}~\bibnamefont {Stander}}, \bibinfo
  {author} {\bibfnamefont {K.}~\bibnamefont {Todd}}, \bibinfo {author}
  {\bibfnamefont {B.}~\bibnamefont {Yang}},\ and\ \bibinfo {author}
  {\bibfnamefont {D.}~\bibnamefont {Goldhaber-Gordon}},\ }\bibfield  {title}
  {\bibinfo {title} {Transport measurements across a tunable potential barrier
  in graphene},\ }\href {https://doi.org/10.1103/PhysRevLett.98.236803}
  {\bibfield  {journal} {\bibinfo  {journal} {Phys. Rev. Lett.}\ }\textbf
  {\bibinfo {volume} {98}},\ \bibinfo {pages} {236803} (\bibinfo {year}
  {2007})}\BibitemShut {NoStop}%
\bibitem [{\citenamefont {Stander}\ \emph {et~al.}(2009)\citenamefont
  {Stander}, \citenamefont {Huard},\ and\ \citenamefont
  {Goldhaber-Gordon}}]{Stander2009}%
  \BibitemOpen
  \bibfield  {author} {\bibinfo {author} {\bibfnamefont {N.}~\bibnamefont
  {Stander}}, \bibinfo {author} {\bibfnamefont {B.}~\bibnamefont {Huard}},\
  and\ \bibinfo {author} {\bibfnamefont {D.}~\bibnamefont {Goldhaber-Gordon}},\
  }\bibfield  {title} {\bibinfo {title} {Evidence for klein tunneling in
  graphene $p\mathrm{\text{\ensuremath{-}}}n$ junctions},\ }\href
  {https://doi.org/10.1103/PhysRevLett.102.026807} {\bibfield  {journal}
  {\bibinfo  {journal} {Phys. Rev. Lett.}\ }\textbf {\bibinfo {volume} {102}},\
  \bibinfo {pages} {026807} (\bibinfo {year} {2009})}\BibitemShut {NoStop}%
\bibitem [{\citenamefont {Sutar}\ \emph {et~al.}(2012)\citenamefont {Sutar},
  \citenamefont {Comfort}, \citenamefont {Liu}, \citenamefont {Taniguchi},
  \citenamefont {Watanabe},\ and\ \citenamefont {Lee}}]{Sutar2012}%
  \BibitemOpen
  \bibfield  {author} {\bibinfo {author} {\bibfnamefont {S.}~\bibnamefont
  {Sutar}}, \bibinfo {author} {\bibfnamefont {E.~S.}\ \bibnamefont {Comfort}},
  \bibinfo {author} {\bibfnamefont {J.}~\bibnamefont {Liu}}, \bibinfo {author}
  {\bibfnamefont {T.}~\bibnamefont {Taniguchi}}, \bibinfo {author}
  {\bibfnamefont {K.}~\bibnamefont {Watanabe}},\ and\ \bibinfo {author}
  {\bibfnamefont {J.~U.}\ \bibnamefont {Lee}},\ }\bibfield  {title} {\bibinfo
  {title} {{Angle-dependent carrier transmission in graphene p-n junctions}},\
  }\href {https://doi.org/10.1021/nl3011897} {\bibfield  {journal} {\bibinfo
  {journal} {Nano Letters}\ }\textbf {\bibinfo {volume} {12}},\ \bibinfo
  {pages} {4460} (\bibinfo {year} {2012})}\BibitemShut {NoStop}%
\bibitem [{\citenamefont {Shytov}\ \emph {et~al.}(2008)\citenamefont {Shytov},
  \citenamefont {Rudner},\ and\ \citenamefont {Levitov}}]{Shytov2008}%
  \BibitemOpen
  \bibfield  {author} {\bibinfo {author} {\bibfnamefont {A.~V.}\ \bibnamefont
  {Shytov}}, \bibinfo {author} {\bibfnamefont {M.~S.}\ \bibnamefont {Rudner}},\
  and\ \bibinfo {author} {\bibfnamefont {L.~S.}\ \bibnamefont {Levitov}},\
  }\bibfield  {title} {\bibinfo {title} {{Klein Backscattering and
  Fabry-P{\'{e}}rot Interference in Graphene Heterojunctions}},\ }\href
  {https://doi.org/10.1103/PhysRevLett.101.156804} {\bibfield  {journal}
  {\bibinfo  {journal} {Physical Review Letters}\ }\textbf {\bibinfo {volume}
  {101}},\ \bibinfo {pages} {156804} (\bibinfo {year} {2008})},\ \Eprint
  {https://arxiv.org/abs/0808.0488} {arXiv:0808.0488} \BibitemShut {NoStop}%
\bibitem [{\citenamefont {Young}\ and\ \citenamefont {Kim}(2009)}]{Young2009}%
  \BibitemOpen
  \bibfield  {author} {\bibinfo {author} {\bibfnamefont {A.~F.}\ \bibnamefont
  {Young}}\ and\ \bibinfo {author} {\bibfnamefont {P.}~\bibnamefont {Kim}},\
  }\bibfield  {title} {\bibinfo {title} {{Quantum interference and Klein
  tunnelling in graphene heterojunctions}},\ }\href
  {https://doi.org/10.1038/nphys1198} {\bibfield  {journal} {\bibinfo
  {journal} {Nature Physics}\ }\textbf {\bibinfo {volume} {5}},\ \bibinfo
  {pages} {222} (\bibinfo {year} {2009})},\ \Eprint
  {https://arxiv.org/abs/0808.0855} {arXiv:0808.0855} \BibitemShut {NoStop}%
\bibitem [{\citenamefont {Rickhaus}\ \emph {et~al.}(2013)\citenamefont
  {Rickhaus}, \citenamefont {Maurand}, \citenamefont {Liu}, \citenamefont
  {Weiss}, \citenamefont {Richter},\ and\ \citenamefont
  {Sch{\"{o}}nenberger}}]{Rickhaus2013}%
  \BibitemOpen
  \bibfield  {author} {\bibinfo {author} {\bibfnamefont {P.}~\bibnamefont
  {Rickhaus}}, \bibinfo {author} {\bibfnamefont {R.}~\bibnamefont {Maurand}},
  \bibinfo {author} {\bibfnamefont {M.~H.}\ \bibnamefont {Liu}}, \bibinfo
  {author} {\bibfnamefont {M.}~\bibnamefont {Weiss}}, \bibinfo {author}
  {\bibfnamefont {K.}~\bibnamefont {Richter}},\ and\ \bibinfo {author}
  {\bibfnamefont {C.}~\bibnamefont {Sch{\"{o}}nenberger}},\ }\bibfield  {title}
  {\bibinfo {title} {{Ballistic interferences in suspended graphene}},\ }\href
  {https://doi.org/10.1038/ncomms3342} {\bibfield  {journal} {\bibinfo
  {journal} {Nature Communications}\ }\textbf {\bibinfo {volume} {4}},\
  \bibinfo {pages} {1} (\bibinfo {year} {2013})},\ \Eprint
  {https://arxiv.org/abs/1304.6590} {arXiv:1304.6590} \BibitemShut {NoStop}%
\bibitem [{\citenamefont {Handschin}\ \emph {et~al.}(2017)\citenamefont
  {Handschin}, \citenamefont {Makk}, \citenamefont {Rickhaus}, \citenamefont
  {Liu}, \citenamefont {Watanabe}, \citenamefont {Taniguchi}, \citenamefont
  {Richter},\ and\ \citenamefont {Sch{\"{o}}nenberger}}]{Handschin2017}%
  \BibitemOpen
  \bibfield  {author} {\bibinfo {author} {\bibfnamefont {C.}~\bibnamefont
  {Handschin}}, \bibinfo {author} {\bibfnamefont {P.}~\bibnamefont {Makk}},
  \bibinfo {author} {\bibfnamefont {P.}~\bibnamefont {Rickhaus}}, \bibinfo
  {author} {\bibfnamefont {M.-H.}\ \bibnamefont {Liu}}, \bibinfo {author}
  {\bibfnamefont {K.}~\bibnamefont {Watanabe}}, \bibinfo {author}
  {\bibfnamefont {T.}~\bibnamefont {Taniguchi}}, \bibinfo {author}
  {\bibfnamefont {K.}~\bibnamefont {Richter}},\ and\ \bibinfo {author}
  {\bibfnamefont {C.}~\bibnamefont {Sch{\"{o}}nenberger}},\ }\bibfield  {title}
  {\bibinfo {title} {{Fabry-P{\'{e}}rot Resonances in a Graphene/hBN
  Moir{\'{e}} Superlattice}},\ }\href
  {https://doi.org/10.1021/acs.nanolett.6b04137} {\bibfield  {journal}
  {\bibinfo  {journal} {Nano Letters}\ }\textbf {\bibinfo {volume} {17}},\
  \bibinfo {pages} {328} (\bibinfo {year} {2017})},\ \Eprint
  {https://arxiv.org/abs/1701.09141} {arXiv:1701.09141} \BibitemShut {NoStop}%
\bibitem [{\citenamefont {Beenakker}(2006)}]{Beenakker:2006}%
  \BibitemOpen
  \bibfield  {author} {\bibinfo {author} {\bibfnamefont {C.~W.~J.}\
  \bibnamefont {Beenakker}},\ }\bibfield  {title} {\bibinfo {title} {Specular
  andreev reflection in graphene},\ }\href
  {https://doi.org/10.1103/PhysRevLett.97.067007} {\bibfield  {journal}
  {\bibinfo  {journal} {Phys. Rev. Lett.}\ }\textbf {\bibinfo {volume} {97}},\
  \bibinfo {pages} {067007} (\bibinfo {year} {2006})}\BibitemShut {NoStop}%
\bibitem [{\citenamefont {Sterpetti}\ \emph {et~al.}(2017)\citenamefont
  {Sterpetti}, \citenamefont {Biscaras}, \citenamefont {Erb},\ and\
  \citenamefont {Shukla}}]{Sterpetti2017}%
  \BibitemOpen
  \bibfield  {author} {\bibinfo {author} {\bibfnamefont {E.}~\bibnamefont
  {Sterpetti}}, \bibinfo {author} {\bibfnamefont {J.}~\bibnamefont {Biscaras}},
  \bibinfo {author} {\bibfnamefont {A.}~\bibnamefont {Erb}},\ and\ \bibinfo
  {author} {\bibfnamefont {A.}~\bibnamefont {Shukla}},\ }\bibfield  {title}
  {\bibinfo {title} {{Comprehensive phase diagram of two-dimensional space
  charge doped Bi$_2$Sr$_2$Ca$_1$Cu$_2$O$_{8+x}$}},\ }\href
  {https://doi.org/10.1038/s41467-017-02104-z} {\bibfield  {journal} {\bibinfo
  {journal} {Nature Communications}\ }\textbf {\bibinfo {volume} {8}},\
  \bibinfo {pages} {2060} (\bibinfo {year} {2017})}\BibitemShut {NoStop}%
\bibitem [{\citenamefont {Mueller}\ \emph {et~al.}(2009)\citenamefont
  {Mueller}, \citenamefont {Xia}, \citenamefont {Freitag}, \citenamefont
  {Tsang},\ and\ \citenamefont {Avouris}}]{Mueller2009}%
  \BibitemOpen
  \bibfield  {author} {\bibinfo {author} {\bibfnamefont {T.}~\bibnamefont
  {Mueller}}, \bibinfo {author} {\bibfnamefont {F.}~\bibnamefont {Xia}},
  \bibinfo {author} {\bibfnamefont {M.}~\bibnamefont {Freitag}}, \bibinfo
  {author} {\bibfnamefont {J.}~\bibnamefont {Tsang}},\ and\ \bibinfo {author}
  {\bibfnamefont {P.}~\bibnamefont {Avouris}},\ }\bibfield  {title} {\bibinfo
  {title} {Role of contacts in graphene transistors: A scanning photocurrent
  study},\ }\href {https://doi.org/10.1103/PhysRevB.79.245430} {\bibfield
  {journal} {\bibinfo  {journal} {Phys. Rev. B}\ }\textbf {\bibinfo {volume}
  {79}},\ \bibinfo {pages} {245430} (\bibinfo {year} {2009})}\BibitemShut
  {NoStop}%
\bibitem [{\citenamefont {Perconte}\ \emph {et~al.}(2018)\citenamefont
  {Perconte}, \citenamefont {Cuellar}, \citenamefont {Moreau-Luchaire},
  \citenamefont {Piquemal-Banci}, \citenamefont {Galceran}, \citenamefont
  {Kidambi}, \citenamefont {Martin}, \citenamefont {Hofmann}, \citenamefont
  {Bernard}, \citenamefont {Dlubak}, \citenamefont {Seneor},\ and\
  \citenamefont {Villegas}}]{Perconte2018}%
  \BibitemOpen
  \bibfield  {author} {\bibinfo {author} {\bibfnamefont {D.}~\bibnamefont
  {Perconte}}, \bibinfo {author} {\bibfnamefont {F.~A.}\ \bibnamefont
  {Cuellar}}, \bibinfo {author} {\bibfnamefont {C.}~\bibnamefont
  {Moreau-Luchaire}}, \bibinfo {author} {\bibfnamefont {M.}~\bibnamefont
  {Piquemal-Banci}}, \bibinfo {author} {\bibfnamefont {R.}~\bibnamefont
  {Galceran}}, \bibinfo {author} {\bibfnamefont {P.~R.}\ \bibnamefont
  {Kidambi}}, \bibinfo {author} {\bibfnamefont {M.~B.}\ \bibnamefont {Martin}},
  \bibinfo {author} {\bibfnamefont {S.}~\bibnamefont {Hofmann}}, \bibinfo
  {author} {\bibfnamefont {R.}~\bibnamefont {Bernard}}, \bibinfo {author}
  {\bibfnamefont {B.}~\bibnamefont {Dlubak}}, \bibinfo {author} {\bibfnamefont
  {P.}~\bibnamefont {Seneor}},\ and\ \bibinfo {author} {\bibfnamefont {J.~E.}\
  \bibnamefont {Villegas}},\ }\bibfield  {title} {\bibinfo {title} {{Tunable
  Klein-like tunnelling of high-temperature superconducting pairs into
  graphene}},\ }\href {https://doi.org/10.1038/NPHYS4278} {\bibfield  {journal}
  {\bibinfo  {journal} {Nature Physics}\ }\textbf {\bibinfo {volume} {14}},\
  \bibinfo {pages} {25} (\bibinfo {year} {2018})}\BibitemShut {NoStop}%
\bibitem [{\citenamefont {Perconte}\ \emph {et~al.}(2020)\citenamefont
  {Perconte}, \citenamefont {Seurre}, \citenamefont {Humbert}, \citenamefont
  {Ulysse}, \citenamefont {Sander}, \citenamefont {Trastoy}, \citenamefont
  {Zatko}, \citenamefont {Godel}, \citenamefont {Kidambi}, \citenamefont
  {Hofmann}, \citenamefont {Zhang}, \citenamefont {Bercioux}, \citenamefont
  {Bergeret}, \citenamefont {Dlubak}, \citenamefont {Seneor},\ and\
  \citenamefont {Villegas}}]{Perconte2020}%
  \BibitemOpen
  \bibfield  {author} {\bibinfo {author} {\bibfnamefont {D.}~\bibnamefont
  {Perconte}}, \bibinfo {author} {\bibfnamefont {K.}~\bibnamefont {Seurre}},
  \bibinfo {author} {\bibfnamefont {V.}~\bibnamefont {Humbert}}, \bibinfo
  {author} {\bibfnamefont {C.}~\bibnamefont {Ulysse}}, \bibinfo {author}
  {\bibfnamefont {A.}~\bibnamefont {Sander}}, \bibinfo {author} {\bibfnamefont
  {J.}~\bibnamefont {Trastoy}}, \bibinfo {author} {\bibfnamefont
  {V.}~\bibnamefont {Zatko}}, \bibinfo {author} {\bibfnamefont
  {F.}~\bibnamefont {Godel}}, \bibinfo {author} {\bibfnamefont {P.~R.}\
  \bibnamefont {Kidambi}}, \bibinfo {author} {\bibfnamefont {S.}~\bibnamefont
  {Hofmann}}, \bibinfo {author} {\bibfnamefont {X.~P.}\ \bibnamefont {Zhang}},
  \bibinfo {author} {\bibfnamefont {D.}~\bibnamefont {Bercioux}}, \bibinfo
  {author} {\bibfnamefont {F.~S.}\ \bibnamefont {Bergeret}}, \bibinfo {author}
  {\bibfnamefont {B.}~\bibnamefont {Dlubak}}, \bibinfo {author} {\bibfnamefont
  {P.}~\bibnamefont {Seneor}},\ and\ \bibinfo {author} {\bibfnamefont {J.~E.}\
  \bibnamefont {Villegas}},\ }\bibfield  {title} {\bibinfo {title} {Long-range
  propagation and interference of $d$-wave superconducting pairs in graphene},\
  }\href {https://doi.org/10.1103/PhysRevLett.125.087002} {\bibfield  {journal}
  {\bibinfo  {journal} {Phys. Rev. Lett.}\ }\textbf {\bibinfo {volume} {125}},\
  \bibinfo {pages} {087002} (\bibinfo {year} {2020})}\BibitemShut {NoStop}%
\bibitem [{\citenamefont {Perconte}\ \emph {et~al.}(2022)\citenamefont
  {Perconte}, \citenamefont {Bercioux}, \citenamefont {Dlubak}, \citenamefont
  {Seneor}, \citenamefont {Bergeret},\ and\ \citenamefont
  {Villegas}}]{Perconte2022}%
  \BibitemOpen
  \bibfield  {author} {\bibinfo {author} {\bibfnamefont {D.}~\bibnamefont
  {Perconte}}, \bibinfo {author} {\bibfnamefont {D.}~\bibnamefont {Bercioux}},
  \bibinfo {author} {\bibfnamefont {B.}~\bibnamefont {Dlubak}}, \bibinfo
  {author} {\bibfnamefont {P.}~\bibnamefont {Seneor}}, \bibinfo {author}
  {\bibfnamefont {F.~S.}\ \bibnamefont {Bergeret}},\ and\ \bibinfo {author}
  {\bibfnamefont {J.~E.}\ \bibnamefont {Villegas}},\ }\bibfield  {title}
  {\bibinfo {title} {{Superconducting Proximity Effect in d ‐Wave
  Cuprate/Graphene Heterostructures}},\ }\href
  {https://doi.org/10.1002/andp.202100559} {\bibfield  {journal} {\bibinfo
  {journal} {Annalen der Physik}\ }\textbf {\bibinfo {volume} {534}},\ \bibinfo
  {pages} {2100559} (\bibinfo {year} {2022})},\ \Eprint
  {https://arxiv.org/abs/2112.09749} {arXiv:2112.09749} \BibitemShut {NoStop}%
\bibitem [{\citenamefont {Linder}\ and\ \citenamefont
  {Sudb\o{}}(2008)}]{Linder2008}%
  \BibitemOpen
  \bibfield  {author} {\bibinfo {author} {\bibfnamefont {J.}~\bibnamefont
  {Linder}}\ and\ \bibinfo {author} {\bibfnamefont {A.}~\bibnamefont
  {Sudb\o{}}},\ }\bibfield  {title} {\bibinfo {title} {{Tunneling conductance
  in $s$- and $d$-wave superconductor-graphene junctions: Extended
  Blonder-Tinkham-Klapwijk formalism}},\ }\href
  {https://doi.org/10.1103/PhysRevB.77.064507} {\bibfield  {journal} {\bibinfo
  {journal} {Phys. Rev. B}\ }\textbf {\bibinfo {volume} {77}},\ \bibinfo
  {pages} {064507} (\bibinfo {year} {2008})}\BibitemShut {NoStop}%
\bibitem [{\citenamefont {Bhattacharjee}\ and\ \citenamefont
  {Sengupta}(2006)}]{Bhattacharjee2006}%
  \BibitemOpen
  \bibfield  {author} {\bibinfo {author} {\bibfnamefont {S.}~\bibnamefont
  {Bhattacharjee}}\ and\ \bibinfo {author} {\bibfnamefont {K.}~\bibnamefont
  {Sengupta}},\ }\bibfield  {title} {\bibinfo {title} {{Tunneling Conductance
  of Graphene NIS Junctions}},\ }\href
  {https://doi.org/10.1103/PhysRevLett.97.217001} {\bibfield  {journal}
  {\bibinfo  {journal} {Phys. Rev. Lett.}\ }\textbf {\bibinfo {volume} {97}},\
  \bibinfo {pages} {217001} (\bibinfo {year} {2006})}\BibitemShut {NoStop}%
\bibitem [{\citenamefont {Sahu}\ \emph {et~al.}(2016)\citenamefont {Sahu},
  \citenamefont {Raychaudhuri},\ and\ \citenamefont {Das}}]{Sahu2016}%
  \BibitemOpen
  \bibfield  {author} {\bibinfo {author} {\bibfnamefont {M.~R.}\ \bibnamefont
  {Sahu}}, \bibinfo {author} {\bibfnamefont {P.}~\bibnamefont {Raychaudhuri}},\
  and\ \bibinfo {author} {\bibfnamefont {A.}~\bibnamefont {Das}},\ }\bibfield
  {title} {\bibinfo {title} {{Andreev reflection near the Dirac point at the
  graphene-NbSe$_2$ junction}},\ }\href
  {https://doi.org/10.1103/PhysRevB.94.235451} {\bibfield  {journal} {\bibinfo
  {journal} {Phys. Rev. B}\ }\textbf {\bibinfo {volume} {94}},\ \bibinfo
  {pages} {235451} (\bibinfo {year} {2016})}\BibitemShut {NoStop}%
\bibitem [{\citenamefont {Dean}\ \emph {et~al.}(2010)\citenamefont {Dean},
  \citenamefont {Young}, \citenamefont {Meric}, \citenamefont {Lee},
  \citenamefont {Wang}, \citenamefont {Sorgenfrei}, \citenamefont {Watanabe},
  \citenamefont {Taniguchi}, \citenamefont {Kim}, \citenamefont {Shepard},\
  and\ \citenamefont {Hone}}]{Dean2010}%
  \BibitemOpen
  \bibfield  {author} {\bibinfo {author} {\bibfnamefont {C.~R.}\ \bibnamefont
  {Dean}}, \bibinfo {author} {\bibfnamefont {A.~F.}\ \bibnamefont {Young}},
  \bibinfo {author} {\bibfnamefont {I.}~\bibnamefont {Meric}}, \bibinfo
  {author} {\bibfnamefont {C.}~\bibnamefont {Lee}}, \bibinfo {author}
  {\bibfnamefont {L.}~\bibnamefont {Wang}}, \bibinfo {author} {\bibfnamefont
  {S.}~\bibnamefont {Sorgenfrei}}, \bibinfo {author} {\bibfnamefont
  {K.}~\bibnamefont {Watanabe}}, \bibinfo {author} {\bibfnamefont
  {T.}~\bibnamefont {Taniguchi}}, \bibinfo {author} {\bibfnamefont
  {P.}~\bibnamefont {Kim}}, \bibinfo {author} {\bibfnamefont {K.~L.}\
  \bibnamefont {Shepard}},\ and\ \bibinfo {author} {\bibfnamefont
  {J.}~\bibnamefont {Hone}},\ }\bibfield  {title} {\bibinfo {title} {{Boron
  nitride substrates for high-quality graphene electronics}},\ }\href
  {https://doi.org/10.1038/nnano.2010.172} {\bibfield  {journal} {\bibinfo
  {journal} {Nature Nanotechnology}\ }\textbf {\bibinfo {volume} {5}},\
  \bibinfo {pages} {722} (\bibinfo {year} {2010})},\ \Eprint
  {https://arxiv.org/abs/1005.4917} {arXiv:1005.4917} \BibitemShut {NoStop}%
\bibitem [{\citenamefont {Sancho}\ \emph {et~al.}(1985)\citenamefont {Sancho},
  \citenamefont {Sancho}, \citenamefont {Sancho},\ and\ \citenamefont
  {Rubio}}]{Sancho1985}%
  \BibitemOpen
  \bibfield  {author} {\bibinfo {author} {\bibfnamefont {M.~P.~L.}\
  \bibnamefont {Sancho}}, \bibinfo {author} {\bibfnamefont {J.~M.~L.}\
  \bibnamefont {Sancho}}, \bibinfo {author} {\bibfnamefont {J.~M.~L.}\
  \bibnamefont {Sancho}},\ and\ \bibinfo {author} {\bibfnamefont
  {J.}~\bibnamefont {Rubio}},\ }\bibfield  {title} {\bibinfo {title} {Highly
  convergent schemes for the calculation of bulk and surface green functions},\
  }\href {https://doi.org/10.1088/0305-4608/15/4/009} {\bibfield  {journal}
  {\bibinfo  {journal} {Journal of Physics F: Metal Physics}\ }\textbf
  {\bibinfo {volume} {15}},\ \bibinfo {pages} {851} (\bibinfo {year}
  {1985})}\BibitemShut {NoStop}%
\bibitem [{\citenamefont {Datta}(1995)}]{Datta1995}%
  \BibitemOpen
  \bibfield  {author} {\bibinfo {author} {\bibfnamefont {S.}~\bibnamefont
  {Datta}},\ }\href {https://doi.org/10.1017/cbo9780511805776} {\emph {\bibinfo
  {title} {Electronic Transport in Mesoscopic Systems}}}\ (\bibinfo
  {publisher} {Cambridge University Press},\ \bibinfo {year}
  {1995})\BibitemShut {NoStop}%
\bibitem [{\citenamefont {Blonder}\ \emph {et~al.}(1982)\citenamefont
  {Blonder}, \citenamefont {Tinkham},\ and\ \citenamefont
  {Klapwijk}}]{Blonder1982}%
  \BibitemOpen
  \bibfield  {author} {\bibinfo {author} {\bibfnamefont {G.~E.}\ \bibnamefont
  {Blonder}}, \bibinfo {author} {\bibfnamefont {M.}~\bibnamefont {Tinkham}},\
  and\ \bibinfo {author} {\bibfnamefont {T.~M.}\ \bibnamefont {Klapwijk}},\
  }\bibfield  {title} {\bibinfo {title} {{Transition from metallic to tunneling
  regimes in superconducting microconstrictions: Excess current, charge
  imbalance, and supercurrent conversion}},\ }\href
  {https://doi.org/10.1103/PhysRevB.25.4515} {\bibfield  {journal} {\bibinfo
  {journal} {Physical Review B}\ }\textbf {\bibinfo {volume} {25}},\ \bibinfo
  {pages} {4515} (\bibinfo {year} {1982})}\BibitemShut {NoStop}%
\bibitem [{\citenamefont {Fisher}\ and\ \citenamefont
  {Lee}(1981)}]{PhysRevB.23.6851}%
  \BibitemOpen
  \bibfield  {author} {\bibinfo {author} {\bibfnamefont {D.~S.}\ \bibnamefont
  {Fisher}}\ and\ \bibinfo {author} {\bibfnamefont {P.~A.}\ \bibnamefont
  {Lee}},\ }\bibfield  {title} {\bibinfo {title} {Relation between conductivity
  and transmission matrix},\ }\href {https://doi.org/10.1103/PhysRevB.23.6851}
  {\bibfield  {journal} {\bibinfo  {journal} {Phys. Rev. B}\ }\textbf {\bibinfo
  {volume} {23}},\ \bibinfo {pages} {6851} (\bibinfo {year}
  {1981})}\BibitemShut {NoStop}%
\bibitem [{\citenamefont {Lado}()}]{Lado}%
  \BibitemOpen
  \bibfield  {author} {\bibinfo {author} {\bibfnamefont {J.~L.}\ \bibnamefont
  {Lado}},\ }\href {https://github.com/joselado/pyqula} {\bibinfo {title}
  {{PyQula}}}\BibitemShut {NoStop}%
\end{thebibliography}%

\end{document}